\documentclass[superscriptaddress,twocolumn,showpacs,preprintnumbers,amsmath,amssymb, pra, reprint, longbibliography, showkeys]{revtex4-1}
\usepackage{amsmath}
\usepackage{amssymb}
\usepackage{graphicx}
\usepackage{stmaryrd}
\usepackage{txfonts}
\usepackage{mathrsfs}
\usepackage{dcolumn}
\usepackage{pbox}
\usepackage{bm}
\usepackage{epsfig}
\usepackage{color}

\usepackage[colorlinks=true, linkcolor=blue, citecolor=blue, urlcolor=blue]{hyperref}
\begin{document}
\preprint{\href{https://doi.org/10.1103/PhysRevB.95.054511}{S.-Z. Lin and V. G. Kogan , Phys. Rev. B {\bf 95}, 054511 (2017).}}

\title{Strain-induced inter-vortex interaction and vortex lattices in tetragonal superconductors}
\author{Shi-Zeng Lin}
\affiliation{Theoretical Division, Los Alamos National Laboratory, Los Alamos, New Mexico 87545, USA}
\author{Vladimir G. Kogan}
\affiliation{Ames Laboratory, DOE and Department of Physics, Iowa State University, Ames, Iowa 50011, USA}

\begin{abstract}
In   superconductors with strong coupling between superconductivity and elasticity  manifested in a strong dependence of transition temperature on pressure, there is an additional contribution to inter-vortex interactions due to the strain field generated by vortices.  When vortex lines are  along the $c$ axis of a  tetragonal crystal, a square vortex lattice (VL) is favored at low vortex densities, because the vortex-induced strains contribution to the inter-vortex interactions is long range. At intermediate magnetic fields, the triangular lattice is stabilized. The triangular lattice evolves to the square lattice upon increasing magnetic field, and eventually the system locks to the square structure. We argue, however, that as magnetic field approaches the upper critical field $H_{c2}$ the elastic inter-vortex interactions disappear faster than the standard London interactions, so that VL should return to the triangular structure. Our results are compared to VLs observed   in the heavy fermion superconductor $\mathrm{CeCoIn_5}$. 

\end{abstract}
%\keywords{strain, vortex configuration, heavy fermion superconductors}

\date{\today}
\maketitle

\section{Introduction}

Vortices are topological excitations in superconductors under magnetic field and they organize in periodic  lattices due to the mutual interaction. 
In isotropic superconductors, the vortices form hexagonal lattices stabilized by the repulsive magnetic interaction. The energy difference between triangular and square lattices is extremely small  \cite{Abrikosov57,PhysRev.133.A1226}, therefore the VLs are sensitive to higher order interaction terms. It was shown that VL evolves  from a triangular  to a square lattice upon increasing magnetic field in tetragonal non-magnetic borocarbides due to nonlocality of the relation between supercurrent and vector potential and the  Fermi surface anisotropy \cite{PhysRevB.55.R8693,PhysRevB.72.060504}, see Ref. \onlinecite{inbook} for a review. The original theories of these structural transitions are based on small nonlocal corrections to the London theory. It was later extended to Ginzburg-Landau approach to investigate the VLs near the upper critical field $H_{c2}$ \cite{PhysRevLett.91.097002}. In another development, thermal fluctuations of vortex positions were shown to substantially change the phase diagram of triangle-to-square transition \cite{PhysRevLett.87.177009}. In particular, these studies have demonstrated an extreme sensitivity of VL structures to small perturbations in inter-vortex interactions. As a result, interpretation of experimentally observed VL structures is extremely difficult, because the inter-vortex interaction depends on crystal anisotropy, pairing symmetry, multi-band characteristics \cite{PhysRevLett.110.087003}, and the possible coexistence of superconductivity with magnetic order \cite{PhysRevB.86.180506}. Despite the decades of effort, the question of VL structures is far from being resolved. To demonstrate that, the example of Nb, the classical type-II superconductor, is  quite illuminating \cite{PhysRevLett.102.136408}. In this work we discuss yet another source which may affect VL structures and even cause the structural transitions: weak elastic crystal perturbations induced by vortices. Square lattice is also favored in models of high-$T_c$ superconductors when both the $s$-  and $d$-wave components are present  \cite{PhysRevB.55.R704}. 

A vortex can perturb the strain field of the crystal that induces additional interactions between vortices \cite{PhysRevB.51.15344,PhysRevB.52.12852,Kogan2013,PhysRevB.87.020503,PhysRevB.53.6682,PhysRevB.68.144515}. In a simple picture, nucleation of the normal vortex core which has a different  density than the surrounding 
 superconductor, induces a strain field. This strain  decays as a power of the distance from the vortex core and mediates long-range interaction between vortices. The strain-induced interaction follows the crystal symmetry. For instance, for vortices directed along the $c$ axis in tetragonal crystal, the strain-induced interaction has four-fold rotational symmetry in the $ab$ plane. It is shown below that because of the long-range nature, the strain-induced interaction, its weakness notwithstanding,  dominates over the short-range magnetic interaction at very small vortex densities (or low magnetic inductions $B$) and at sufficiently high vortex densities where the elastic part of the free energy increases as $B^2$, whereas the standard contribution of inter-vortex interactions to the free energy scales as $B$. 
 
 The coupling between superconductivity and elasticity is characterized by the rate of change of the critical  temperature $T_c$  with respect to stress/pressure $p$, i.e. by derivatives $d T_c/d p$. 
It was argued in Ref.\,\cite{PhysRevB.51.15344} that in NbSe$_2$ with  $d T_c/d p\approx 0.5\,$K/GPa, the magneto-elastic interactions might be responsible for observed VL structures. In a heavy-fermion superconductor $\mathrm{CeCoIn_5}$, $d T_c/d p\approx 0.3\,$K/GPa. \cite{0953-8984-16-49-008} 
 For iron-based materials, $d T_c/d p$ is on the order of K/GPa and varies with doping. In some of those, e.g. in $\mathrm{Ca(Fe_{1-x}Co_x)_2As_2}$, $d T_c/d p \approx -60\,$K/GPa  which is by two orders of magnitude larger than common values \cite{PhysRevB.86.220511}. Hence, all these materials are good candidates  for observing the vortex structure evolution and transitions  caused by strain induced interactions.

The elastic contribution to intervortex interaction in tetragonal materials has been discussed in Ref.\,\onlinecite{Kogan2013}. However, the VL structures in the presence of the new interaction have not been studied. The present work aims to fill this gap. We will also discuss the possible relevance of strain-induced interaction to the VL transitions observed by small angle neutron scattering in $\mathrm{CeCoIn_5}$   \cite{bianchi_superconducting_2008,white_observations_2010,PhysRevLett.108.087002}.

%%%%%%%
\section{Model}
%%%%%%%

Within our model, the total   energy density   $ F $ associated with the VL consists of the superconducting contribution $ F_{m}$ and the elastic energy density  
\begin{equation}\label{eq1}
    F_{e}=\lambda_{iklm}u_{ik}u_{lm}/2,
\end{equation}
where $\lambda_{iklm}$ are elastic moduli and $u_{ik}$ are strains.  Summation over repeated indices is implied throughout the paper. We will focus on  tetragonal crystals. For brevity we denote the six independent moduli in the crystal frame $(a,b,c)$ as: $\lambda_{aaaa}=\lambda_{bbbb}=\lambda_1$, $\lambda_{aabb}=\lambda_2$, $\lambda_{abab}=\lambda_3$, $\lambda_{cccc}=\lambda_4$, $\lambda_{aacc}=\lambda_{bbcc}=\lambda_5$ and $\lambda_{acac}=\lambda_{bcbc}=\lambda_6$   \cite{LandauBookElasticity}. We do not use the common two-indices notation of elastic moduli because of the symmetry restrictions. The transformations of the $4$th rank tensor $\lambda_{iklm}$ are more transparent in the form adopted in Ref. \onlinecite{LandauBookElasticity}.

Within the London approximation, we have for the magnetic and kinetic parts of the superconducting free energy density \cite{PhysRevB.24.1572,PhysRevB.38.2439}: 
\begin{equation}\label{eq2}
   8\pi F _m= {\bm h}^2+\lambda_L^2 m_{ik} (\nabla\times {\bm h})_i(\nabla\times{\bm h})_k .
\end{equation}
where $\bm h $ is the local magnetic field, the mass tensor $m_{ik}$ accounts for the uniaxial anisotropy, and $\lambda_L$ is the geometric average of the penetration depths.

We take the Bardeen-Stephen approximation for the vortex core as a normal region of size of $\xi$  \cite{PhysRev.140.A1197}. The crystal expands or shrinks in the normal region: $(V_n-V_s)/V_s=H_c \partial_p H_c/4\pi$, where $V_{n,s}$ are the specific volumes of the normal and superconducting  phases, $p$ is the pressure, and $H_c$ is the thermodynamic critical field \cite{LandauBookEM}. One can consider the vortex as a point (line) source in two (three) dimensions, which induces the strain field \cite{PhysRevB.51.15344}. This London-type model is, of course, oversimplified and works far from the upper critical field $H_{c2}(T)$ and for   $\lambda_L\gg \xi$.

Let vortex lines be directed along the unit vector $\bm l=(\cos\varphi\sin\theta,\ \sin\varphi\sin\theta,\ \cos\theta)$ in the tetragonal crystal frame ($\theta$ is the angle between $\bm c$ and $\bm l$, $\varphi$ is the angle between the $a$ axis and the  projection of $\bm l$ onto the $ab$ plane). We introduce also the ``vortex frame"  $(X,\ Y,\ Z)$ such that the  $Z$ direction is along the vortex line. A vector $ \bm{V}$ in the rotated frame is related to the vector $ \bm{v}$ in the crystal frame by a rotation $ \bm{v}=\hat{ \bm{O}} \bm{V}$, with
\begin{equation}\label{eq3}
    {\bf{\hat O}} = \left( {\begin{array}{*{20}{c}}
{ - \sin   \varphi}&{ - \cos  \theta  \cos  \varphi  }&{\cos   \varphi  \sin   \theta   }\\
{\cos   \varphi  }&{ - \cos  \theta  \sin \varphi   }&{\sin   \theta  \sin   \varphi  }\\
0&{\sin   \theta  }&{\cos  \theta   }
\end{array}} \right).
\end{equation}
The strain tensor $u_{ik}$ in the crystal frame is related to that $U_{\alpha\beta}$, with $\alpha,\ \beta=X,\ Y,\ Z$, in the rotated frame according to
\begin{equation}
u_{ik}=\frac{1}{2}\left( {{O_{i\alpha }}{O_{k\beta }} + {O_{k\alpha }}{O_{i\beta }}} \right){U_{\alpha \beta }},
\end{equation}
and the elastic moduli $\Lambda_{\alpha\beta\gamma\eta}$ in the rotated frame are 
\begin{equation}\label{eqLambda}
{\Lambda _{\alpha \beta \gamma \eta }} \equiv \frac{\lambda _{iklm}}{4}\left( {{O_{i\alpha }}{O_{k\beta }} + {O_{k\alpha }}{O_{i\beta }}} \right)\left( {{O_{l\gamma }}{O_{m\eta }} + {O_{m\gamma }}{O_{l\eta }}} \right).
\end{equation}
The stress tensor $\sigma_{\alpha\beta}$ is
\begin{equation}\label{eq4}
    {\sigma _{\alpha \beta }} = \partial { F_e}/\partial {U_{\alpha \beta }} = {\Lambda _{\alpha \beta \gamma \eta }}{U_{\gamma \eta }}.
\end{equation}

It is argued in Ref.\,\cite{Kogan2013} that for the vortex orientation $\bm l$ along the principal crystal directions, the problem of elastic perturbation caused by straight vortices can be considered as {\it planar}, i.e. the deformations $\bm U\perp\bm l$  everywhere  and the strains  $U_{\alpha Z}=0$. The elasticity problem then simplifies considerably, and below we consider two vortex orientations: $\bm l\parallel \bm c$ and $\bm l\perp\bm c$. In the first case the vortex frame coincides with the crystal frame, the corresponding elastic moduli are listed above.
The elastic moduli in the second case are compiled in  Appendix \ref{AppedixA}. In both cases, we have ${\Lambda _{XXXY}} = {\Lambda _{XXYX}} = {\Lambda _{YYXY}} = {\Lambda _{YYYYX}} = {\Lambda _{ZZXY}} = {\Lambda _{ZZYX}} = 0$. 

As always in planar problems, the components of the stress tensor are not independent \cite{LandauBookElasticity}. After a simple algebra, one can exclude $U_{\gamma\eta}$ 
from Eqs.\,\eqref{eq4} to obtain:
\begin{eqnarray}
 \sigma _{ZZ} &=& \left(\frac{ D_2 }{d} - 1\right) \sigma _{XX} + \left(\frac{ D_1} {d} - 1\right) \sigma _{YY} \,,\label{eq6}\\
 D_1&=&d+ \Lambda _{ZZYY}  \Lambda _{XXXX}  -  \Lambda _{ZZXX} \Lambda _{XXYY}  \,, \label{DX}\\
 D_2&=&d+ \Lambda _{ZZXX} \Lambda _{YYYY}  -  \Lambda _{ZZYY} \Lambda _{XXYY} \,,\label{DY}\\
   d&=&\Lambda _{XXXX} \Lambda _{YYYY}  - \Lambda _{XXYY}^2 \,.\label{d}
\end{eqnarray}
Equilibrium conditions $\partial {\sigma _{\alpha \beta }}/\partial {X_\beta } = 0$ read:
\begin{equation}
\frac{{\partial {\sigma _{XX}}}}{{\partial X}} + \frac{{\partial {\sigma _{XY}}}}{{\partial Y}} = 0,\ \ \ \frac{{\partial {\sigma _{YX}}}}{{\partial X}} + \frac{{\partial {\sigma _{YY}}}}{{\partial Y}} = 0\,.
\end{equation}
The solution can be written as 
\begin{equation}
{\sigma _{XX}} = \frac{{{\partial ^2}\chi }}{{\partial {Y^2}}},\ \ {\sigma _{YY}} = \frac{{{\partial ^2}\chi }}{{\partial {X^2}}},\ \ {\sigma _{XY}} =  - \frac{{{\partial ^2}\chi }}{{\partial X\partial Y}},
\end{equation}
with an arbitrary function $\chi(X, Y)$ \cite{LandauBookElasticity}. 

Using the condition $\sigma_{\alpha\alpha}=-3 p$, we obtain  for $\chi(X, Y)$:
\begin{equation}\label{eq8}
\frac{D_1}{d}\frac{{{\partial ^2}\chi }}{{\partial {X^2}}}+\frac{D_2}{d}\frac{{{\partial ^2}\chi }}{{\partial {Y^2}}}=-3p.
\end{equation}
To calculate the stress field induced by a vortex in otherwise unrestrained sample, we note that the pressure $p=0$ while the vortex can be considered as a singular source of the stress field \cite{Kogan2013}. 
Because the stress field is long-ranged, we can approximate the source term using a delta function, $2\pi S_0\delta(\bm{R}-\bm{R}_v)$ with $\bm{R}_v=(X_v,\ Y_v)$ being the vortex position \cite{Kogan2013}.
Equation \eqref{eq8} with a delta-source can be solved by the two-dimensional Fourier transform, and both $\sigma_{\alpha\beta}(\bm{R}_v, \bm{k})$ and $U_{\alpha\beta}(\bm{R}_v, \bm{k})$ can be calculated
\begin{align}
\begin{split}
{\sigma _{XX}}(\mathbf{k}) = \frac{{d2\pi {S_0}k_Y^2}}{{{D_1}k_X^2 + {D_2}k_Y^2}},\ {\sigma _{YY}}(\mathbf{k}) = \frac{{d2\pi {S_0}k_X^2}}{{{D_1}k_X^2 + {D_2}k_Y^2}},\\
{\sigma _{XY}}(\mathbf{k}) =  - \frac{{d2\pi {S_0}{k_X}{k_Y}}}{{{D_1}k_X^2 + {D_2}k_Y^2}},\\
{U_{YX}}(\mathbf{k}) = \frac{{ - \pi d{S_0}{k_X}{k_Y}}}{{({D_1}k_X^2 + {D_2}k_Y^2){\Lambda _{XYXY}}}},\\
{U_{XX}}(\mathbf{k}) = \frac{{2\pi {S_0}({\Lambda _{YYYY}}k_Y^2 - {\Lambda _{XXYY}}k_X^2)}}{{{D_1}k_X^2 + {D_2}k_Y^2}},\\
{U_{YY}}(\mathbf{k}) = \frac{{2\pi {S_0}({\Lambda _{XXXX}}k_X^2 - {\Lambda _{XXYY}}k_Y^2)}}{{{D_1}k_X^2 + {D_2}k_Y^2}}.
\end{split}
\end{align}

The elastic contribution to the interaction energy (per unit length) of a vortex at the origin and another one at $\bm{R}_v$ is
\begin{equation}\label{eq9}
 {\cal E}_e(\bm{R}_v) = \int \frac{ {\bm d}^2  {\bm k}} { 4 \pi ^2 }{\sigma _{\alpha \beta }}\left( {0,{\bm{k}}} \right){U_{\alpha \beta }}\left( {\bm{R_v}, - {\bm{k}}} \right).
\end{equation}
For VL along the $c$ axis, $\theta=0$, $D_1=D_2=D$, the strain induced interaction becomes
\begin{align}
{{\cal E}_e(\bm{R}_v)}=\left(\frac{S_0}{D}\right)^2d\left(\lambda_1+\lambda_2-\frac{d}{2\lambda_3}\right)\int d^2 \bm{k}\,\frac{k_X^4+k_Y^4}{ k ^4} e^{-i \bm{k}\cdot \bm{R}_v}\nonumber\\
=\left(\frac{S_0}{D}\right)^2\frac{\pi d}{R_v^2\lambda_3}{\left( {{\lambda _1} + {\lambda _2}} \right)\left( {2{\lambda _3} - {\lambda _1} + {\lambda _2}} \right)}\cos(4\phi),\label{eq10}
\end{align}
where the azimuth of the second vortex position is $\phi=\tan^{-1}(Y_v/X_v)$. The interaction decays as $1/R_v^2$ and has four-fold rotational symmetry. When $2\lambda_3-\lambda_1+\lambda_2>0$, the elastic interaction changes from repulsion at $\phi=0$ to attraction at $\phi=\pi/4$. There is no angular independent contribution in Eq. \eqref{eq10}, indicating the absence of elastic contribution to vortex-vortex interactions in isotropic superconductors. For instance, it was shown that vortices along the $c$-axis of hexagonal crystals do not interact elastically \cite{PhysRevB.51.15344}.

For VL along the $ab$ plane, the straightforward algebra results in
\begin{align}
{{\cal E}_e( \bm{R}_v)}= {{{\left( {\frac{d S_0}{D_2}} \right)}^2}\int } d^2 \bm{k} e^ { - {i}{\bm{k}}\cdot{{\bm{R}}_v}}\frac{{{f_X}{{\left( {{D_1}D_2^{ - 1}k_X^2} \right)}^2} + {f_Y}k_Y^4}}{{{{\left( {{D_1}D_2^{ - 1}k_X^2 + k_Y^2} \right)}^2}}}\nonumber\\
=\left( {\frac{d S_0}{D_2}} \right)^2\frac{{\pi \sqrt {{D_2}D_1^{ - 1}} }}{{X_v^2{D_2}D_1^{ - 1} + Y_v^2}}[f_+\cos(4\phi')+2f_-\cos(2\phi')],\label{eq11}
\end{align}
where
\begin{eqnarray}
f_X={\frac{{{\Lambda _{XXXX}}}}{d} - \left( {\frac{1}{{{\Lambda _{XYXY}}}} - \frac{{2{\Lambda _{XXYY}}}}{d}} \right)\frac{{{D_2}}}{{2{D_1}}}} , \\
f_Y={\frac{{{\Lambda _{YYYY}}}}{d} - \left( {\frac{1}{{{\Lambda _{XYXY}}}} - \frac{{2{\Lambda _{XXYY}}}}{d}} \right)\frac{{{D_2}}}{{2{D_1}}}},\\ 
f_\pm=f_Y\pm f_X\,,\quad  \phi'=\tan^{-1} \left( {\sqrt {{D_1}D_2^{ - 1}} {Y_v}/{X_v}} \right) \,. 
\end{eqnarray}
This interaction  has the two-fold rotational symmetry.  The strain induced interaction depends on the vortex orientation through $\Lambda_{\alpha\beta\gamma\eta}(\varphi)$.

%%%%%%
\section{Vortex Lattices}
%%%%%%%%%

By minimizing the total interaction energy we obtain the equilibrium VL configuration. For vortices along the $c$ axis, the lattice unit cell   is a rhombus. We consider two vortex configurations with the rhombus diagonal    either along [100] or [110] directions. The unit cell vectors of the reciprocal lattice  with the rhombus diagonal in [100] are
\begin{equation}
 \bm{G}_{1,2}=\frac{{2\pi }}{{a\sin  {\beta} }}\left[\sin\left(\frac{\beta}{2}\right)\hat x \pm \cos\left(\frac{\beta}{2}\right)\hat y\right]\,.
\end{equation}
For the rhombus diagonal in [110],  we have 
\begin{equation}
{{\bm{G}}_{1,2}} = \frac{{2\pi }}{{a\sin  \beta }}\left[\sin \left( {\frac{\beta }{2} \mp \frac{\pi }{4}} \right)\hat x \pm \cos \left( {\frac{\beta }{2} \mp \frac{\pi }{4}} \right)\hat y\right],
\end{equation}
where $\beta$ is the apex angle and $\hat{x}$, $\hat{y}$ are unit vectors along the crystal directions $a,b$. The length   $a=\sqrt{\Phi_0/B\sin\beta}$ relates to the VL size in real space, where $\Phi_0$ is the flux quantum and $B$ is the magnetic induction. 

\begin{figure}[b]
	\psfig{figure=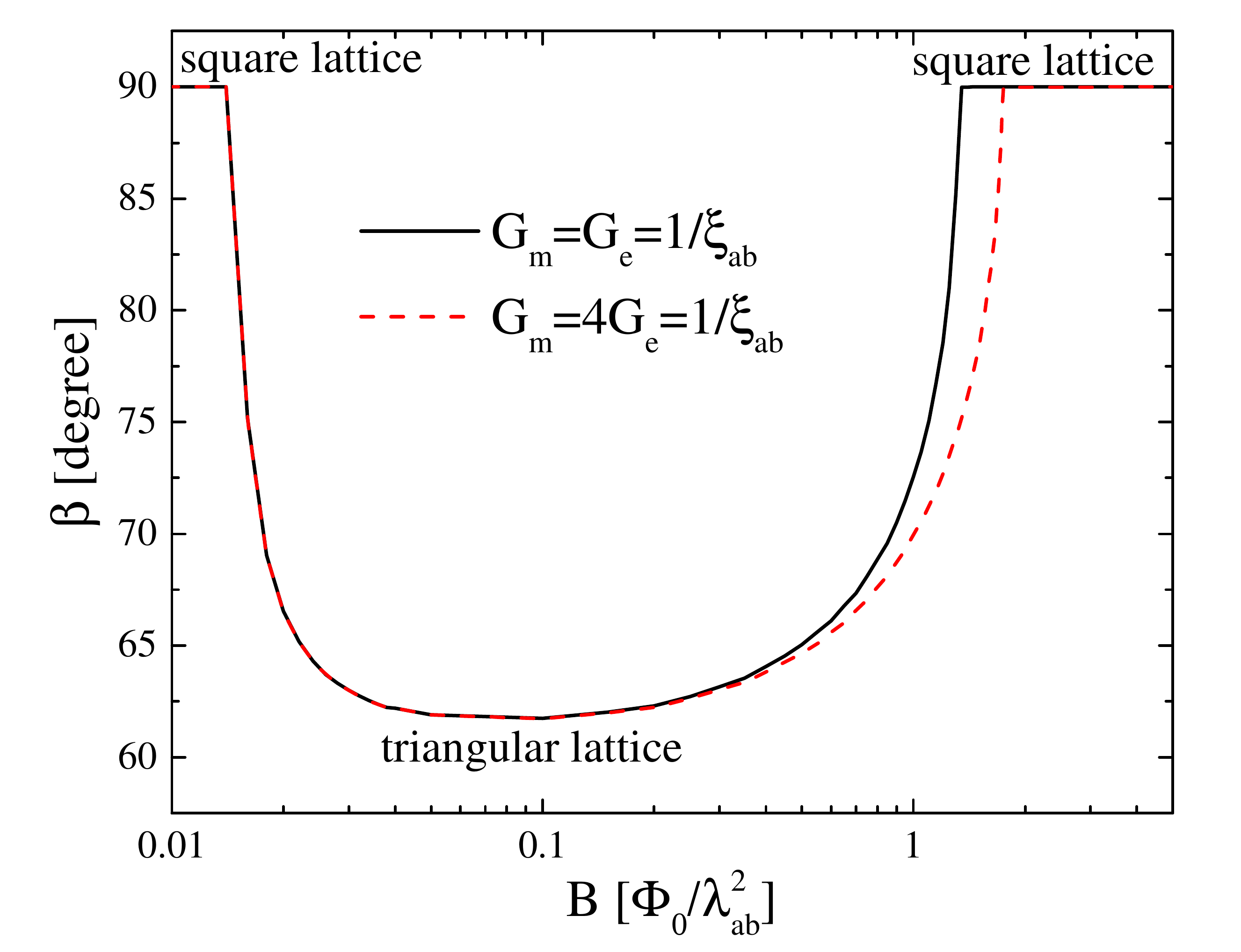,width=\columnwidth}
	\caption{(color online) The equilibrium apex angle $\beta$ of the rhombic unit cell for vortex lines  along   $c$. Vortices form a square lattice both at low and high fields, and the VL is triangular  at intermediate fields. For comparison, the results for two different $G_e$'s are displayed. The field at the transition from the triangular to square VL at higher field increases when $G_e$ is reduced. 
	} \label{f2}
\end{figure}

The free energy density can be expressed as a sum over the reciprocal lattice, see e.g. \cite{PhysRevB.38.2439}: 
\begin{equation}
F=\frac{B^2}{8\pi}\sum_{\mathbf{G}\neq 0}\left(\frac{e^ { - G ^2 / G_{m} ^2 } }{{1 + { G}^2 \lambda_{ab}^2}} +\eta \, \frac{G_x^4 + G_y^4}{  G^4  } e^ { -   G ^2 / G_e ^2 }  \right),  \label{eq12}
\end{equation}
where we introduced two cutoffs, $G_{m}$ and $G_{e}$, for the magnetic and elastic contributions in divergent sums. Meanwhile we have excluded the contribution from $\mathbf{G}=0$ component. The magnetic contribution at $\mathbf{G}=0$ is the magnetic static energy for a uniform magnetic field, which does not determine the profile of vortex lattice. The elastic contribution diverges at $\mathbf{G}=0$, which is unphysical. This divergence is avoided by the strain produced by external pressure, in analogy to the requirement of charge neutrality in electrostatic problem. Here, the factor 
 \begin{equation}
 \eta = \frac{16\pi^3 S_0^2 d}{\Phi_0^2 D^2}\left(\lambda_1+\lambda_2-\frac{d}{2\lambda_3}\right),
 \end{equation} 
characterizes the strain contribution to the inter-vortex interaction. Here $\lambda_{ab}$ and $\lambda_{c}$, $\xi_{ab}$, $\xi_c$ discussed below are the anisotropic London penetration depth and superconducting coherence length respectively. 

We roughly estimate $\eta\sim S_0^2/\Phi_0^2\tilde{\lambda}$, where $\tilde\lambda$ is the order of magnitude of elastic constants. Here $S_0\sim \tilde{\lambda}\xi_{ab}^2 \frac{H_c^2}{T_c}\frac{d T_c}{d p}\left(\ln\frac{\lambda_{ab}}{\xi_{ab}}\right)^2$. For $dT_c/d p\approx 1\ \mathrm{K/GPa}$, $H_c=1\ \mathrm{T}$ and $\tilde{\lambda}\sim 10^{12}\ \mathrm{erg/cm^3}$, we obtain $\eta\sim 5\times 10^{-4}$ \cite{Kogan2013}. 

 \begin{figure}[t]
\psfig{figure=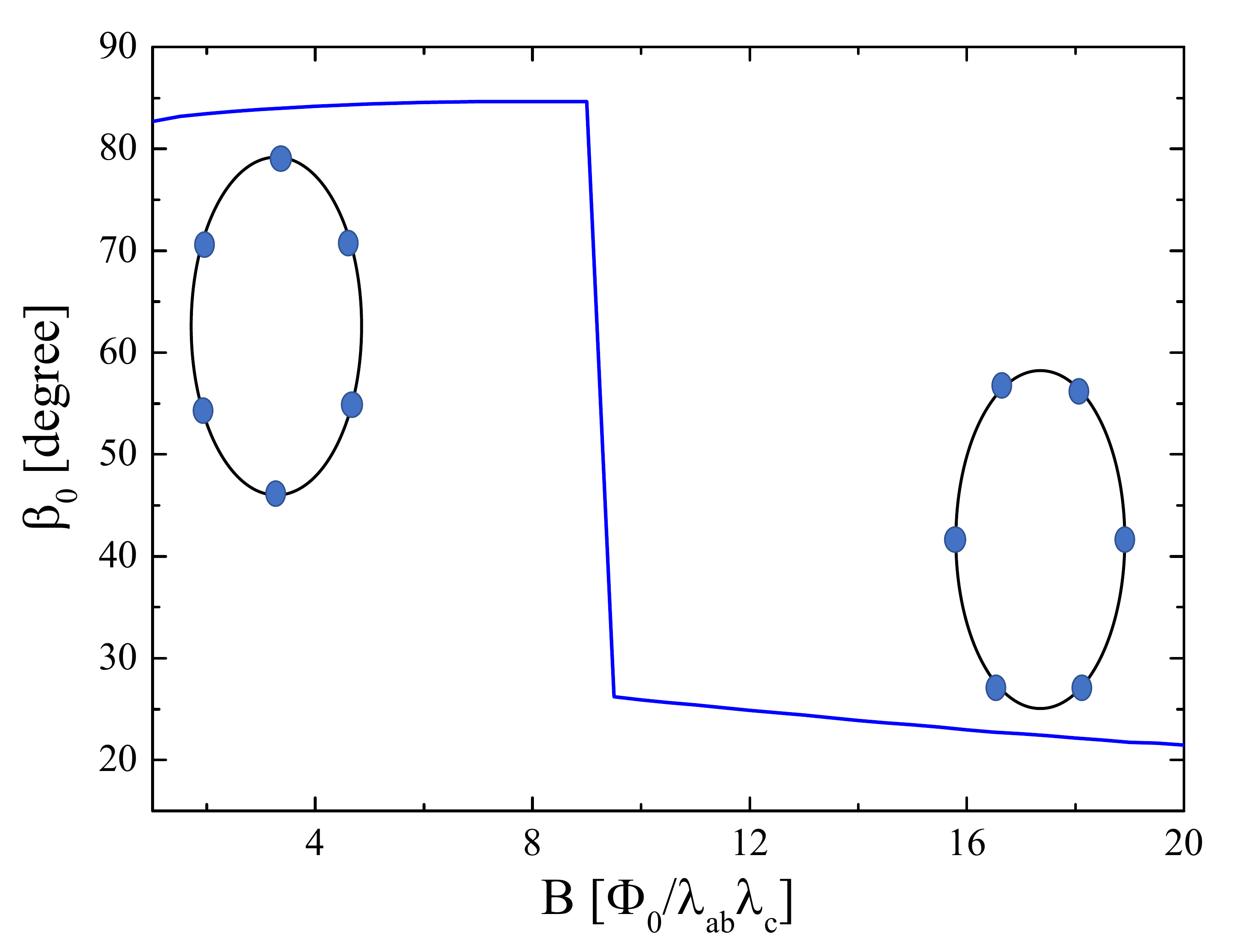,width=\columnwidth}
\caption{(color online) The apex angle $\beta$ of the rhombic unit cell for $\bm B\perp \bm c$. Insets are the sketches of the corresponding Bragg peaks of maximum intensity in the  momentum space at low and high fields. Here $\eta_X=0$, $\eta_Y=0.002$, $\bar\gamma=2$, $\gamma=2$ and $\bar\kappa=30$.
} \label{f3}

\end{figure}
 \begin{figure}[b]
\psfig{figure=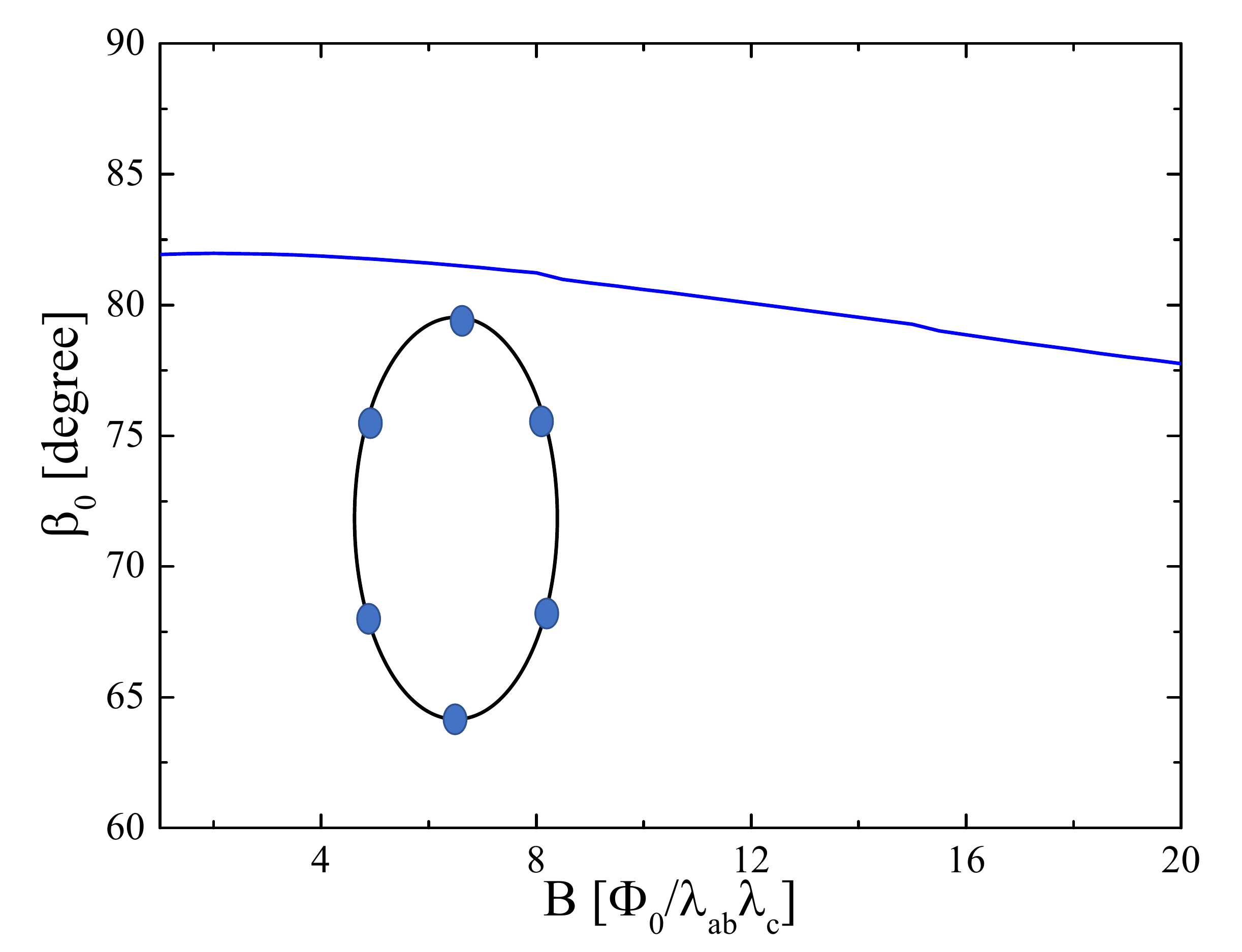,width=\columnwidth}
\caption{(color online) The apex angle $\beta$ of the rhombic unit cell when for $\bm B\perp \bm c$. Inset  is a sketch of the corresponding Bragg peaks of maximum intensity in the  momentum space. Here $\eta_X=0.001$, $\eta_Y=0$, $\bar\gamma=2$, $\gamma=2$ and $\bar\kappa=30$.
} \label{f4}
\end{figure}
  
 We perform numerical summation in Eq.\,\eqref{eq12} to obtain  $\beta$ corresponding to minimum energy for a given magnetic field. We find that the rhombus with diagonal along [100] has lower energy when $\eta>0$. The results for $\eta=0.005$ is shown in Fig.\,\ref{f2}. At low fields where the separation between vortices is larger than $\lambda_{ab}$, the long-range strain induced interaction is dominant, therefore the square VL is stabilized. In intermediate fields, the magnetic interaction favors a triangular VL. The triangular VL then evolves continuously to a square lattice for a high vortex density, because the long-range elastic interaction is $\propto B^2$, whereas the large field London interaction energy goes as $\Phi_0B/\lambda_L^2$. Hence, at large $B$ the elastic contribution is dominant and the VL follows the crystal symmetry. 
For a larger $\eta$, the intermediate region for triangular structures VL shrinks, and eventually disappears for a sufficiently strong elastic interaction. 

We have introduced a cutoff $G_m=1/\xi_{ab}$ for $G$ to exclude distances shorter than the core size $\xi_{ab}$ (or $G>1/\xi_{ab}$) by adding a damping factor $\exp(-G^2\xi_{ab}^2)$ in Eq.\,\eqref{eq12}, the standard procedure in the London approximation. 

Since the source of 
the strain generated by vortex can be related not only to the vortex core, but to the supercurrents around it \cite{PhysRevB.68.144515}, the cutoff $G_e$ in the elastic part of energy \eqref{eq12} can differ from the London cutoff $G_m$.  
Numerical results and the threshold field to stabilize the square lattice depend on the cutoff $G_e$ as shown in Fig.\,\ref{f2}. Nevertheless, the qualitative feature that the square VL is favored by the strain induced interaction is robust against the cutoff.  

 As is seen from Eq.\,\eqref{eq10}, the VL configuration for a negative $\eta$ is related to the corresponding positive $\eta$ by rotation of the whole lattice by $\pi/4$. Therefore in this case the diagonal of the rhombic unit cell is along the [110].

Next, we consider   vortex lines   in the $ab$ plane. Because of the anisotropic penetration depth, the VL is no longer  hexagonal   in the absence of strain. Taking the crystal anisotropy into account, the London contribution to the energy density  is
\begin{equation}
F_m = \frac{{{B^2}}}{{8\pi }} \sum_{G\neq 0} \frac{{\exp \left( { - G_X^2\xi _{ab}^2 - G_Y^2\xi _c^2} \right)}}{{1 + G_Y^2\lambda _{ab}^2 + G_X^2\lambda _c^2}},
\end{equation}
and the contribution due to strain is
\begin{equation}
 F_e = 2\left(\frac{\pi B d S_0}{\Phi_0 D_2}\right)^2\sum_{G\neq 0} \frac{{{f_X}(D_1D_2^{-1}G_X)^4 + {f_Y}G_Y^4}}{[(D_1D_2^{-1}G_X)^2+G_Y^2]^2} e^{ { - G_X^2\xi _{ab}^2 - G_Y^2\xi _c^2}}.
\end{equation}
We rescale the length $G_Y\lambda_{ab}\rightarrow G_Y$ and $G_X\lambda_{c}\rightarrow G_X$, such that ${F}_{m}$ becomes isotropic. The total energy density  then reads as
\begin{equation}
 F  = \frac{B^2}{8\pi}  \sum_{ \bm{G}\neq 0} e^{  { - { \bm{G}^2}/{{\bar\kappa} ^2}} }\left[ \frac{1}{{1 + { \bm{G}^2}}} + \frac{{{\eta_X}{{\left( {G_X^2{\bar\gamma}^{-2}} \right)}^2} + {\eta_Y}G_Y^4}}{{{{\left( {G_Y^2 + G_X^2{\bar\gamma}^{-2}} \right)}^2}}}\right] \qquad\qquad
 \label{eq23}
\end{equation}
with $\bar\kappa=\lambda_{ab}/\xi_c=\lambda_{c}/\xi_{ab}$ and ${\bar\gamma}^{-2}=D_1D_2^{-1}\lambda_{ab}^2/\lambda_c^2$ \cite{note1}.

To check possible VL structure transitions, we   take 
\begin{equation}
\eta_{X,Y}\equiv 16\pi^3\left(\frac{d S_0}{\Phi_0 D_2}\right)^2 f_{X,Y},
\end{equation}
as free parameters to obtain the equilibrium vortex configurations. We consider the VL   rhombic unit cell with the diagonal along the $X$ axis in the rescaled frame. The apex angle is $\beta$. The   apex angle  $\beta_0$, before rescaling   is given by $\tan(\beta_0/2)=\tan(\beta/2)/\gamma$ with $\gamma=\lambda_c/\lambda_{ab}$. Below we present the results for two typical parameters.

The equilibrium apex angle $\beta_0$ for $\eta_X=0$, $\eta_Y=0.002$, $\gamma=2$ and $\bar\gamma=2$ is shown in Fig. \ref{f3}. At low fields $\beta_0\approx 83^\circ$ and it drops to about $\beta_0\approx 24^\circ$ at high fields. The jump indicates a reorientation of the VL upon increasing field. To relate to the neutron scattering measurements, we depict in the inset the reciprocal unit vector in the first shell for both VL orientations.  This reorientation resembles the one observed in $\mathrm{CeCoIn_5}$ for field along the [110] direction.   

The results for 
$\eta_X=0.001$, $\eta_Y=0$, $\gamma=2$  and $\bar\gamma=2$ are displayed in Fig. \ref{f4}. The apex angle depends weakly on the field and there is no reorientation, similar to behavior  observed in $\mathrm{CeCoIn_5}$ for field along the [100] direction. We note that since $\eta_X$ and $\eta_Y$ depend on the field angle through $\Lambda_{\alpha\beta\gamma\eta}(\varphi)$, it is possible that VL reorients when field rotates in the $ab$ plane.

%%%%%%%%
\section{Discussion}
%%%%%%%%%%
We expect   stabilization of the square VL and the reorientation of VLs due to vortex-induced  strain  to occur in a broad class of materials and in heavy fermion superconductors, in particular. 

In heavy fermion superconductors, such as $\mathrm{CeCoIn_5}$ and $\mathrm{CeRhIn_5}$, $T_c$ depends strongly on pressure, pointing to  a possible   strain-induced inter-vortex interactions which affect the VL structures.  For $\mathrm{CeCoIn_5}$, $\partial T_c/\partial p\approx 0.3\ \mathrm{K/GPa}$ at the ambient pressure. In a large family of Fe-based materials, this derivative is larger yet and depends on doping; in some of them $\partial T_c/\partial p$ can be one or two orders of magnitude larger \cite{PhysRevB.86.220511}. 

The small angle neutron scattering data on VLs in $\mathrm{CeCoIn_5}$ are available 
\cite{bianchi_superconducting_2008,white_observations_2010,PhysRevLett.108.087002}. 
 At low fields along the $c$ axis and low temperatures, the VL is triangular (rhombic). Upon increasing field, the VL becomes square. With further increase of the magnetic field, the VL becomes triangular again.
 
One possible explanation to the triangular-square VL transition in $\mathrm{CeCoIn_5}$ could be the strain-induced intervortex interaction. It is worth noting that strain-induced interactions  are not the only possible mechanism to stabilize the square lattice. It may also be due to non-local corrections to the London interactions due to the basic nonlocality of current-field relation in superconductors, as has been demonstrated theoretically and experimentally for borocarbides \cite{PhysRevB.55.R8693}. However, the high-field square-to-triangle transition cannot be explained by the nonlocal effects. It can  be caused by fluctuations of vortices near the upper critical field $H_{c2}$ \cite{PhysRevLett.87.177009} or by the strong Pauli pair breaking \cite{PhysRevLett.101.027001}. 

Our London-type model is inapplicable near $H_{c2}$. Physically, near $H_{c2}$ the system is nearly uniform and there are no inhomogeneities to cause elastic perturbations. Therefore, one can argue that within the magneto-elastic scenario considered here,  the vortex induced strains disappear faster than the standard inter-vortex interaction when the field increases toward $H_{c2}(T)$. The strain-induced interaction has been estimated in \cite{PhysRevB.51.15344}:  
  \begin{equation}
 F_e\sim{\tilde\lambda}\left(\frac{\Phi_0B }{16\pi^2\lambda_L^2T_c}   \frac{\partial T_c}{\partial p} \right)^2\,. 
\end{equation}
The inter-vortex interaction contribution to the London free energy density in intermediate fields is
  \begin{equation}
 F_m\sim \frac{\Phi_0B }{32\pi^2\lambda_L^2 } \ln\left(\frac{H_{c2}}{B}\right). 
\end{equation}
As $T$ increases toward $T_c(B)$ at a fixed $B$, $F_e \propto 1/\lambda_L^4$ decreases faster than $F_m \propto 1/\lambda_L^2$. As a result VL favors the triangular lattice because of the dominant magnetic interaction. 
Hence, both the  triangle-to-square evolution of VLs and the square-to-triangle transition on approach to $H_{c2}$ can, in principle, be attributed to the existence and variation of the vortex induced strains. 

Experimentally,  for $\mathrm{CeCoIn_5}$ in the field parallel to  [110], the VL rotates near $B\approx 8\ \mathrm{T}$, similar to that shown in Fig.\,\ref{f3}. For field along [100], the observed VL deforms weakly, which is akin to behavior in Fig.\,\ref{f4}. Unfortunately,  direct comparison between the theory and experiment is not possible at the moment because the elastic moduli of $\mathrm{CeCoIn_5}$ are not known.

To summarize, we have studied   vortex lattice configurations in tetragonal superconductors   taking into account the strain field created by vortices. When vortex lines are directed along the $c$ axis and for a weak vortex-strain coupling, the square vortex lattice is stabilized both at high and low vortex densities, while the triangular vortex lattice is favored at  intermediate densities. In the presence of a strong vortex-strain coupling, the square vortex lattice may be favored in the whole field region. When vortex lines lie in the $ab$ plane, the vortex lattice can reorient with increasing magnetic field. Our results are in qualitative agreement with  the vortex evolution and  transitions in $\mathrm{CeCoIn_5}$.

\section{Acknowledgements}
The authors thank Lev  Bulaevskii, Roman Movshovich, Leonardo Civale, Duk Y. Kim, and Ian Fisher  for helpful discussions. The work by SZL was carried out under the auspices of the U.S. DOE Contract No. DE-AC52-06NA25396 through the LDRD program. VGK was supported by the U.S. Department of Energy, Office of Science, Basic Energy Sciences, Materials
Sciences and Engineering Division. The Ames Laboratory is operated for the U.S. DOE by Iowa State University under Contract No. DE-AC02-07CH11358.

\appendix

\section{Elastic moduli in the rotated frame}\label{AppedixA}

With the help of Eq. \eqref{eqLambda}
 we evaluate the elastic moduli in the vortex frame   when the vortex axis $Z$ is   in the $ab$ plane of a tetragonal crystal. The azimuthal angle $\varphi$ is the angle between the $Z$ and $a$ axes and the axis $Y$ coincides with the $c$ axis. 
\begin{eqnarray}
\Lambda_{XXXX}&=&[3\lambda_1+\lambda_2+2\lambda_3+(\lambda_1-\lambda_2-2\lambda_3)\cos(4\varphi)]/4,\nonumber\\
\Lambda_{XXXZ}&=&(\lambda_1-\lambda_2-2\lambda_3)\sin(4\varphi)/4,\nonumber\\
\Lambda_{XXZZ}&=&[\lambda_1+3\lambda_2-2\lambda_3+(-\lambda_1+\lambda_2+2\lambda_3)\cos(4\varphi)]/4,\nonumber\\
\Lambda_{XZXZ}&=&[\lambda_1-\lambda_2+2\lambda_3+(-\lambda_1+\lambda_2+2\lambda_3)\cos(4\varphi)]/4,\nonumber\\
\Lambda_{XZZZ}&=&(-\lambda_1+\lambda_2+2\lambda_3)\sin(4\varphi)/4,\nonumber
\end{eqnarray}
and $\Lambda_{XXYY}=\Lambda_{YYZZ}=\lambda_5$, $\Lambda_{YYYY}=\lambda_4$, $\Lambda_{XYXY}=\Lambda_{YZYZ}=\lambda_6$.

\bibliography{reference}

\begin{thebibliography}{29}
\bibitem [{not()}]{note1}%
  \BibitemOpen
  \href@noop {} {}\bibinfo {note} {The equality
  $\lambda_{ab}/\xi_c=\lambda_{c}/\xi_{ab}$ holds only if the anisotropies of
  the penetration depth and of the coherence length are the same $
  \lambda_{ab}/\lambda_{c}=\xi_{ab}/\xi_c$, which is in general true only at
  $T_c$. At low $T$'s, this anisotropies may differ substantially; e.g., for
  MgB$_2$ at $T=0$, $\gamma_\lambda=\lambda_{c}/ \lambda_{ab}\approx 1$,
  whereas $\gamma_\xi=\xi_{ab}/\xi_c\approx 6$. Still, we keep
  $\gamma_\lambda=\gamma_\xi$ not to make our formulas too
  cumbersome.}\BibitemShut {Stop}%
\end{thebibliography}

\end{document}